\documentclass[a4paper,12pt]{article}

\usepackage{graphics}
\usepackage{latexsym}
\usepackage{amsmath}
\usepackage{amssymb}
\newcommand{\sectiono}[1]{\section{#1}\setcounter{equation}{0}}

\begin{document}
\begin{titlepage}
\thispagestyle{empty}
\begin{flushright}
UT-03-29\\
hep-th/0309063\\
September, 2003 
\end{flushright}

\vskip 1.5 cm

\begin{center}
\noindent{\textbf{\LARGE{ Tadpole Cancellation  \\\vspace{0.5cm}
 in Unoriented Liouville Theory 
\vspace{0.5cm}\\
}}} 
\vskip 1.5cm
\noindent{\large{Yu Nakayama}\footnote{E-mail: nakayama@hep-th.phys.s.u-tokyo.ac.jp}}\\ 
\vspace{1cm}
\noindent{\small{\textit{Department of Physics, Faculty of Science, University of 
Tokyo}} \\ \vspace{2mm}
\small{\textit{Hongo 7-3-1, Bunkyo-ku, Tokyo 113-0033, Japan}}}
\end{center}
\vspace{1cm}
\begin{abstract}
The tadpole cancellation in the unoriented Liouville theory is discussed. Using two different methods --- the free field method and the boundary-crosscap state method, we derive one-loop divergences. Both methods require two D1-branes with the symplectic gauge group to cancel the orientifold tadpole divergence. However, the finite part left is different in each method and this difference is studied. We also discuss the validity of the free field method and the possible applications of our result.
\end{abstract}

\end{titlepage}
\newpage
\baselineskip 6mm


\sectiono{Introduction}\label{sec:introduction}
The matrix-Liouville correspondence is one of the simplest examples of the open-closed duality\footnote{For the older review, see e.g.\cite{Seiberg:1990eb} \cite{Ginsparg:is} \cite{Klebanov:1991qa} \cite{Martinec:1991kn}.}. The new interpretation of the matrix model \cite{McGreevy:2003kb}\cite{McGreevy:2003ep}\cite{Klebanov:2003km}\cite{Douglas:2003up} is that we regard this matrix quantum mechanics as the effective dynamics of the open tachyon filed which lives on D0-branes stacked at the boundary of the Liouville direction $\phi \to \infty$ \cite{Zamolodchikov:2001ah}.

Extending this idea to the unoriented Liouville theory (or with an orientifold plane) is interesting, since we less understand the open-closed duality with an orientifold plane than the case without one. However, it is well-known that unoriented theories usually have tadpoles when the space-filling orientifold plane is introduced. Since the appearance of tadpoles implies divergences in many amplitudes (usually referred to as an internal inconsistency), it must be cancelled somehow. Typically introducing space-filling D-branes saves the situation \cite{Green:sg}\cite{Ohta:nq}\cite{Polchinski:1987tu}. In this short note, we would like to examine this possibility in the unoriented Liouville theory.

The organization of this paper is as follows. In section 2, we use the so called ``free field method" to calculate the Liouville partition functions on the cylinder, Klein bottle and M\"obius strip. We cancel the orientifold tadpole completely by introducing two D1-branes with $Sp(2)$ group. In section 3, we use the boundary-crosscap state method to recalculate the tadpole in the unoriented Liouville theory, and we find that introducing two D1-branes with $Sp(2)$ gauge group cancels the tadpole divergence. The implication of the finite part left is discussed. In section 4, we conclude the paper and discuss some connections with the dual matrix model calculation.

Note added: after the original version of this note was submitted to the arxive, S.~Hirano \cite{Hirano:2003}\cite{Hirano:2003au} pointed out that the second sign in the crosscap one-point function (\ref{eq:RP2}) derived in \cite{Hikida:2002bt} is wrong, namely the previous $\cosh(2\pi p)- 1$ should be $\cosh(2\pi p) + 1$. This sign error changes the crosscap-boundary state method calculation considerably. We modify our argument as little as possible to be consistent.

\sectiono{Naive Cancellation --- Free Field Method}\label{sec:nc}

The action of the Liouville theory is 
\begin{equation}
S = \int d^2z\sqrt{g} \left(\frac{1}{4\pi}g^{ab}\partial_a\phi \partial_b \phi + \frac{1}{4\pi}QR\phi + \mu e^{2b\phi}\right),
\end{equation}
where $Q=b+b^{-1}$, $1+ 6Q^2 = c$ and in the 2 dimensional string case, $b=1$. Since it is not free, the Liouville partition function or any correlation function is difficult to calculate. However, it is proposed that some special correlation functions may be calculated by the free field perturbative path integral except the zero mode contribution \cite{Gupta:fu}. 

The logic is as follows. The correlation function of the Liouville theory satisfies an exact Ward-Takahashi identity,
\begin{equation}
\langle e^{2\alpha_1 \phi} e^{2\alpha_2\phi} \cdots e^{2\alpha_N\phi} \rangle \propto \mu^{\frac{(1-g)Q-\sum_i \alpha_i}{b}}, \label{eq:ewt}
\end{equation}
where $g$ is a world-sheet genus, and this determines the exact dependence of $\mu$ in any correlation function. This dependence clearly shows that the perturbative treatment of $\mu$ breaks down in general, but it may be possible when the power of $\mu$ is an integer. The conjecture is when the power of $\mu$ is an integer, the perturbative calculation is valid\footnote{This reminds us of the supersymmetric instanton calculation, where, in general, the instanton calculation does not yield the right charge conservation, but when it does, it gives the exact answer.}. After integrating over the zero mode of $\phi$, the calculation left is just the Coulomb gas free field path integral. Note that the Coulomb gas perturbative calculation yields a nonzero answer just when the power of $\mu$ is an integer.

The simplest example of this calculation is the torus partition function. Since the power of $\mu$ should be zero (up to a logarithmic renormalization term, see the actual form below), there is no need of $\mu$ perturbation. That means we can calculate the partition function as if $\mu$ were zero, except for the zero mode integration. The zero mode integration yields the Liouville volume:
\begin{equation}
V_\phi = \int d\phi e^{-\mu e^{2b\phi}} = -\frac{1}{2b} \log\mu.
\end{equation}
The rest of the path integral is free,
\begin{equation}
Z_{T^2} = V_\phi V_X\int_\mathcal{F}d^2\tau \frac{|\eta(q)|^4}{2\tau_2} \frac{1}{(2\pi\sqrt{\tau_2})^2|\eta(q)|^4} = - V_X\frac{1}{48\pi}\log\mu,
\end{equation}
where $\mathcal{F}$ is the fundamental domain of the torus moduli, and $\eta(q) = \eta(e^{2\pi i\tau}) $ is the Dedekind eta function. The first $|\eta(q)|^4$ comes from the ghost oscillator, and $1/(2\tau_2)$ comes from the Beltrami differential. $1/(2\pi\sqrt{\tau_2}|\eta(q)|^2)$ is the contribution from the non-zero Liouville mode and from that of the free $X$ field\footnote{ For a precise coefficient see e.g. Polchinski's text book \cite{Polchinski:rq}. Our convention here is $\alpha'=1$ and $d^2\tau = d\tau_1d\tau_2$. }. This gives the exact partition function which matches the matrix model computation.

There are other amplitudes which can be calculated in this way. For instance, the three point function ``on mass-shell" (which means $ s = (Q- \sum_i \alpha_i)/b$ is an integer) calculated in this perturbative approach yields the residue of the DOZZ formula \cite{Dorn:1994xn}\cite{Zamolodchikov:1995aa} which is supposed to be exact for any $s$. 

Then it is natural to consider the Klein bottle partition function in the same manner because it has the same $\mu$ dependence and the world sheet curvature is also zero. The free field path integration results in,
\begin{equation}
Z_{K_2} = V_\phi V_X \int_0^\infty \frac{dt}{4t} \frac{1}{4\pi^2 t},
\end{equation}
which diverges when $t \to 0$. If we do the (formal) modular transformation $ s = \pi/2t$, it becomes more clear
\begin{equation}
Z_{K_2} = V_\phi V_X \int_0^\infty \frac{ds}{8\pi^3}.
\end{equation}
This is nothing but the massless tadpole amplitude.

As in the ordinary string theory, the massless tadpole should be cancelled by other D-branes. For this purpose, we introduce the cylinder and the M\"obius strip partition function as follows ($n$ is a number of D1-branes and $+,-$ sign corresponds to $SO(2n)$ and $Sp(2n)$ respectively):
\begin{equation}
Z_{C_2} = n^2V_\phi V_X  \int_0^\infty \frac{dt}{4t}\frac{1}{8\pi^2t} = n^2 V_\phi V_X  \int_0^\infty \frac{ds}{8\pi^3 4},
\end{equation}
\begin{equation}
Z_{M_2} = \pm nV_\phi V_X  \int_0^\infty \frac{dt}{4t}\frac{1}{8\pi^2t} = \pm n V_\phi V_X  \int_0^\infty \frac{ds}{8\pi^3 }.
\end{equation}
In the cylinder case, the modular transformation\footnote{Because these partition functions are divergent, the modular transformation is rather formal. We do a kind of ``dimensional regularization" to determine their conventional modular transformation (see \cite{Polchinski:rq}) which is needed to cancel the tadpole.} is $s = \pi/t$, and in the M\"obius strip case, it is $s = \pi/4t$. Combining these amplitudes, in order to cancel the tadpole divergence, we should take $n=2$ and $Sp(2)$ gauge group.

This is just the same tadpole cancellation in the free 2D string (without a ``tachyon background"). There might be some questions. What kind of D1-branes did we use?
 As is discussed in \cite{Fateev:2000ik}, D1-branes in the Liouville theory have a continuous parameter $s$ which is related to the boundary cosmological constant $\mu_B$ as 
 \begin{equation}
 \cosh^2\pi bs = \frac{\mu_B^2}{\mu} \sin\pi b^2.
 \end{equation}
We treat D1-branes such that on which strings have the Neumann boundary condition. Therefore, we might guess $\mu_B = 0$ on these D1-branes. Is this interpretation consistent with the disk one-point function? What happens if we turn on the boundary cosmological constant? We answer these questions after investigating the boundary-crosscap state formalism in the next section.

\sectiono{Boundary-Crosscap State Method}\label{sec:bc}
As in the usual string theory, the one-loop partition function can be calculated in two ways: either via an open string loop diagram or via a closed string exchange diagram. In section 2 under the assumption of the free field method, we computed the free open loop diagrams and modular transformed them to the closed exchange diagrams. These closed exchange diagrams are also essentially free. Although there might be some nontrivial cancellation because they are integrated, it is quite natural to suppose the boundary-crosscap states are also those of the free field theory. 
However, this cannot be true, for we know there is a nontrivial one-point function on the disk and on the projective plane. The disk one-point function \cite{Fateev:2000ik} is
\begin{equation}
\Psi_s(\nu) = \frac{2^{-1/4}\Gamma(1+2ib\nu)\Gamma(1+2ib^{-1}\nu)\cos(2\pi s \nu)}{-2i\pi \nu} (\pi\mu\gamma(b^2))^{-i\nu/b},
\end{equation}
where $\alpha = \frac{Q}{2} +i\nu$. This defines the boundary state as a linear combination of the Ishibashi states
\begin{equation}
\langle B_s| = \int_{-\infty}^\infty d\nu \Psi_s(\nu)\langle B, \nu|
\end{equation}
where the Ishibashi states \cite{Ishibashi:1988kg} are defined as
\begin{equation}
\langle B,P| q^{H}|B,P'\rangle = \delta(P-P') \frac{q^{P^2}}{\eta(q)}.
\end{equation}

On the other hand, the one-point function on the projective plane \cite{Hikida:2002bt}\footnote{We thank S.~Hirano and O.~Bergman \cite{Hirano:2003} for pointing out the sign error in this one-point function.} is
\begin{eqnarray}
\Psi_C(\nu) &=& \frac{2^{-3/4}\Gamma(1+2ib\nu)\Gamma(1+2ib^{-1}\nu)}{-2\pi i\nu}(\pi\mu\gamma(b^2))^{-i\nu/b} \times \cr
&\times& \left(\cosh\left(\pi\nu(b+b^{-1})\right) + \cosh\left(\pi\nu(b-b^{-1})\right)\right), \label{eq:RP2}
\end{eqnarray}
which defines the crosscap state as 
\begin{equation}
\langle C| = \int_{-\infty}^\infty d\nu \Psi_C(\nu) \langle C,P| ,
\end{equation}
where the crosscap Ishibashi states \cite{Ishibashi:1988kg} are defined by
\begin{eqnarray}
\langle C,P| q^{H}|C,P'\rangle = \delta(P-P') \frac{q^{P^2}}{\eta(q)}, \cr
\langle B,P| q^{H}|C,P'\rangle = \delta(P-P') \frac{q^{P^2/2}}{\eta(-\sqrt{q})}.
\end{eqnarray}

First, we calculate the Klein bottle partition function as follows,
\begin{equation}
Z_{K_2} =V_X \int_0^\infty  d\tau \left(\int_{-\infty}^{\infty} d\nu \Psi_C(\nu)\Psi_C(-\nu) \frac{q^{\nu^2}}{\eta(q)}\right)\frac{2}{\sqrt{2}\eta(q)}\eta(q)^2 ,
\end{equation}
where $q = e^{-2\pi\tau}$ which is the closed channel modular parameter, and because of the explicit momentum integration we do not have the Liouville volume factor here. Substituting (\ref{eq:RP2}) and setting $b=1$, we obtain
\begin{equation}
Z_{K_2} = V_X \int_{-\infty}^\infty  \frac{d\nu}{4\pi} \frac{[\cosh(2\pi\nu)+1]^2}{[\sinh(2\pi\nu)]^2\nu^2}.
\end{equation}
This partition function is ultraviolet finite but infrared divergent. This divergence shows the existence of the massless tadpole.

To cancel the tadpole, we should calculate other Euler number zero partition functions. The cylinder partition function \cite{Martinec:2003ka} is (with $n$ D1-branes whose boundary cosmological constants are same for simplicity)
\begin{eqnarray}
Z_{C_2} &=& n^2V_X \int_{0}^\infty  d\tau \left(\int_{-\infty}^{\infty} d\nu \Psi_s(\nu)\Psi_s(-\nu) \frac{q^{\nu^2}}{\eta(q)}\right)\frac{1}{\sqrt{2}\eta(q)}\eta(q)^2 \cr
&=& n^2 V_X \int_{-\infty}^\infty \frac{d\nu}{4\pi} \frac{[\cos(2\pi s\nu)]^2}{[\sinh(2\pi\nu)]^2\nu^2}, \label{eq:anu}
\end{eqnarray}
which diverges as $\nu \to 0$. This is a closed channel infrared divergence, i.e. a tadpole divergence. 

The M\"obius strip partition function is\footnote{The correspondence between the $\pm$ here and the gauge group choice (orientifold operation on the Chan-Paton indices) is not so obvious. To fix this, consider the open sector calculation $\pm n \mathrm{Tr}_{O} [\Omega e^{-2\pi tH}]$. If we take the limit $t \to \infty$, this contribution is positive for the $SO(n)$ and negative for $Sp(n)$. Then we modular transform this and compare it with the boundary-crosscap state calculation in the $s = \pi/4t \to 0$ limit. The relevant sign is determined by the $\lim_{\nu \to 0} \Psi_c(\nu)\Psi_s(-\nu)$ which is positive in our normalization of $\Psi_c(\nu)$ and $\Psi_s(\nu)$. Thus $+$,$-$ indeed corresponds to $SO(n)$ and $Sp(n)$ respectively.}
\begin{equation}
Z_{M_2} = \pm n V_X \int_{-\infty}^\infty \frac{d\nu}{2\pi} \frac{[\cos(2\pi s\nu)][\cosh(2\pi\nu)+1]}{[\sinh(2\pi\nu)]^2\nu^2},
\end{equation}
which also diverges as $\nu \to 0$. Combining all these partition functions we obtain,
\begin{equation}
Z_{1loop} = V_X \int_{-\infty}^\infty \frac{d\nu}{4\pi} \frac{[\cosh(2\pi\nu)+1  \pm n \cos(2\pi s\nu)]^2}{[\sinh(2\pi\nu)]^2\nu^2}.
\end{equation}
The infrared tadpole divergence can be cancelled if we choose the $Sp(2)$ gauge group for two D1-branes, irrespective of the value of $s$.

This is the main result of this section. Using the boundary-crosscap state formalism, we obtained the same tadpole cancellation condition i.e. two D1-branes with the gauge group $Sp(2)$. However, the finite part left is different in each methods. In the free field calculation, the partition function completely vanishes. On the other hand, in the boundary-crosscap calculation, there is a finite part left even if we choose $s$ to be $i\frac{\pi}{2}$, which corresponds to the $\mu_B=0$. It is interesting to see whether this finite part can be seen from the matrix quantum mechanics point of view. For the unoriented Liouville theory, the corresponding dual matrix quantum mechanics should be $SO(2N)$ or $Sp(2N)$ quantum mechanics ``living on the D0-branes"\footnote{Actually the tadpole cancelled matrix quantum mechanics should choose $Sp(2N)$ gauge group. Once we choose the normalization of the crosscap state to be minus that of the D1-brane so as to cancel the tadpole, the D0-brane gauge field should also be $Sp(2N)$ \cite{Hirano:2003au}. Intuitive argument in the matrix model point of view is that, for $Sp$ theory, a twisted loop has an extra minus sign and this is necessary to cancel the fundamental loop.}, which is also considered as a discretized version of the 2D unoriented quantum gravity with a boson on it. It is important to note that this matrix quantum mechanics is finite. Therefore, the matrix quantum mechanics predicts that the Klein bottle partition function should be finite. In contrast, the matrix model loop amplitude reproduces that of the divergent boundary-boundary annular diagram (\ref{eq:anu}) as was studied in \cite{Martinec:2003ka}. Therefore, the need for the tadpole cancellation from the matrix model point of view is rather puzzling and further study is needed \cite{Hirano:2003}.

Then, how can we explain the small mismatch between the two methods? The free boundary-crosscap condition seems to be wrong. However, we cannot totally distrust the free field method described in section 2, for the free field method was actually used to derive these boundary-crosscap sate, the DOZZ three point function formula etc. We will discuss this further in the next section.

\sectiono{Summary and Discussion}\label{sec:summary}

In this short note, the tadpole cancellation of the unoriented Liouville theory was discussed. By the free field method, it was shown that there is a tadpole in the Klein bottle partition function, and it must be cancelled by introducing two D1-branes. On the other hand, by the boundary-crosscap state method, it was shown that the tadpole divergence can be cancelled, but there is a finite part left after the cancellation of the tadpole infrared divergence. Especially whether or not the Klein bottle partition function can be calculated by the free field method is a problem.

At this point, we should note that we actually used the free field method to derive the boundary-crosscap state. For example, the fusion coefficient of the degenerate primary operator $V_{-b/2} = e^{-b\phi}$ can be written as
\begin{equation}
V_{-b/2} V_{\alpha} = C_{+} [V_{\alpha-b/2}] + C_{-}[V_{\alpha+b/2}],
\end{equation}
where $C_{-}$ was actually derived by the free field perturbation method \cite{Teschner:1995yf}\cite{Fateev:2000ik}. Since it requires just the first order insertion of the Liouville interaction, it was calculated by the free field Coulomb gas integral:
\begin{equation}
C_{-} = -\mu \int d^2z \langle V_{\alpha}(0) V_{-b/2} e^{2b\phi(z)}V_{Q-\alpha-b/2}(\infty)\rangle_{Q,free}.
\end{equation}
Thus, it may be necessary to reexamine the condition when the free field method can be reliably used. 

Let us discuss the physical implications of our result, supposing that the boundary-crosscap method is correct. The first thing is the divergence of the disk partition function. This means that the boundary Liouville theory is not a suitable string background as a one-loop corrected string theory (as opposed to the closed Liouville theory which is conformal and becomes a correct string background in any order of the string perturbation, as we know that it is dual to the matrix model which yields the all genus computation.). When this happens in the infrared limit of the closed channel, it is the Fischler-Suskind mechanism \cite{Fischler:ci}\cite{Fischler:tb} that saves the situation. We hope the similar background deformation might work and we suppose this has some effects on the ``rolling tachyon" computation beyond the tree level \cite{Sen:2002nu}\cite{Lambert:2003zr}\cite{Karczmarek:2003xm}. 

To apply the unoriented partition function to the rolling tachyon computation, we simply replace the $X$ free Neumann boundary condition with that of the rolling tachyon boundary state $|B_t\rangle$. For concreteness, we take the half s-brane (rolling) state which corresponds to the boundary interaction $\lambda e^{X^0}$. The imaginary part of the one-loop partition function is related to the emitted closed string number and energy via the optical theorem:
\begin{equation}
N = 2\mathrm{Im}Z= \int_0^\infty \frac{dp}{2p} |A_{p}|^2,
\end{equation}
where $A(p)$ is an on-shell decaying amplitude. We consider here the rolling D0-brane in the unoriented tadpole cancelled background with two D1-branes. This case is also interesting because in the unoriented Liouville theory, the decaying D0-branes can be regarded as a rolling fermion from the top of the inverse harmonic oscillator potential in the matrix model point of view \cite{McGreevy:2003kb}\cite{McGreevy:2003ep}\cite{Klebanov:2003km}. As was discussed in the last section, the cancelled one-loop partition function is finite, and there is no imaginary part,
\begin{equation}
N_{1loop} = 2\mathrm{Im} (Z_{K_2} + Z_{C_2} + Z_{M_2})= 0.
\end{equation}
The contribution from the D0-brane cylinder partition function was computed in the literature \cite{Sen:2002nu}\cite{McGreevy:2003kb}\cite{McGreevy:2003ep}\cite{Klebanov:2003km}. We need the following matrix elements:
\begin{eqnarray}
\langle Z_{1,1} | p \rangle &=& c_1\sinh(2\pi p) \cr
\langle B_t | E \rangle &=&  c_2\frac{1}{\sinh(\pi E)} 
\end{eqnarray}
where $c_1$ and $c_2$ are irrelevant energy dependent phase factors. Note in our convention, the mass shell condition is $E = 2p$. The decaying rate is logarithmically (ultraviolet and infrared) divergent
\begin{equation}
N_{C2} = \int_0^\infty \frac{dp}{2p} (const)^2.
\end{equation}
Under this background, there is also an interference term:
\begin{equation}
N_{interference} = c\int_{-\infty}^\infty\frac{dp}{2p} 2\pi\delta(p)\frac{(\cosh(2\pi p)+1-\cos(2\pi s p))}{\sinh(2\pi p)} \frac{\sinh(2\pi p)}{\sinh(2\pi p)} = finite,
\end{equation}
which is finite. The emitted energy as closed strings is zero, which can be obtained replacing $dp/p$ with $dp$. Since finitely many zero energy strings are emitted, the emitted energy is obviously zero.

The cylinder divergence is just an artefact of our approximation, for the emitted energy cannot exceed the energy of the decaying D0-brane. Actually, without an orientifold plane, the interpretation of the D0-brane as a rolling fermion in the matrix model enables us to obtain the full string decaying amplitude which is finite as it should be. Thus, it is interesting to see whether there is such an interpretation of the decaying D0-brane in a tadpole cancelled orientifold plane background which gives a finite answer and reproduces the above finite correction from the interference diagram.

The second point we would like to discuss is the effect of supersymmetry \cite{Takayanagi:2003sm}\cite{Douglas:2003up}\cite{McGreevy:2003dn}. Though we have concentrated only on the bosonic Liouville theory in this paper, an extension to the supersymmetric case is intriguing. As is discussed in \cite{Douglas:2003up}, the type 0B theory has a D1-brane which has R-R tadpole. Whether this tadpole can be cancelled by introducing a space-filling orientifold plane is an interesting question. Another interesting model is the space-time supersymmetric Liouville theory which was studied in \cite{McGreevy:2003dn}. Being space-time supersymmetric, divergences may be eliminated at least at the partition function level. 

The last point we would like to discuss is the possibility of the topological string interpretation of the unoriented Liouville theory compactified on the self-dual radius. As is well-known, the ordinary oriented Liouville theory is dual to the topological closed B model on the deformed conifold \cite{Ghoshal:1995wm}. This has a following duality chain \cite{Gopakumar:1998ki}\cite{Ooguri:2002gx}\cite{Aganagic:2002wv}. This closed B model on the deformed conifold is large $N$ dual to the open B model on the resolved conifold, which is, by the mirror symmetry, equivalent to the nonperturbative part of the open topological A model on the deformed conifold. Finally, the nonperturbative part of the open topological A model on the deformed conifold is the nonperturbative part of the $U(N)$ Chern-Simons theory on $S^3$, which is found to be the volume factor of $U(N)$ group. In a word, the Liouville partition function at the self dual radius is just the volume of the $U(N)$ group! We naturally expect the volume factor of $SO(2N)$ or $Sp(2N)$ has some relations with the unoriented Liouville theory. However, the closer look at the $SO(2N)$ or $Sp(2N)$ group volume shows that the partition functions with two cross-caps (like the Klein bottle partition function) are always zero \cite{Sinha:2000ap}\cite{Ooguri:2002gx}. Hence, the orientifold tadpole must be cancelled in the theory if any noncritical string theory is the dual of it\footnote{Since we do not have the Liouville volume factor in the partition function calculated by the boundary-crosscap method, the one-loop partition function is not proportional to the logarithm of $\mu$. This might suggest the absence of the two crosscap contribution to the partition function, for this constant part can be attributed to the multiplicative renormalization of $\mu$. However, there is an unsolved subtlety here: in fact we are able to do the same calculation in the $c=0$ pure 2D gravity case, and we find that the Klein bottle partition function calculated by the crosscap state method is divergent but do not depend on $\mu$. On the contrary, the $SO(2N)$ matrix model predicts $\log\mu$ contribution in the Klein bottle partition function \cite{Brezin:1990xr}\cite{Harris:1990kc}. Curiously, this can be correctly obtained by the free field method. Therefore a good regularization scheme is needed to compare the boundary-crosscap state calculation to the matrix model result.}.


\section*{Acknowledgements}
This work was motivated by the informal seminar held at the university of Tokyo
high energy physics theory group. The author would like to thank all the participants in the seminar. The author also acknowledges valuable discussions with T.~Eguchi, Y.~Hikida, Y.~Tachikawa and T.~Takayanagi.

The author would thank S.~Hirano for pointing out the sign error and offering other valuable comments.

\nopagebreak


\end{document}